\documentclass{iau} 
\usepackage{graphicx}

\title[IAU-S315~~Gas and star formation from z=3 to 0] %% give here short title %%
{The Evolution of Gas and Star Formation from z=3 to z=0}

\author[F. Combes and PHIBSS coll. ]   %% give here short author list %%
{Francoise Combes$^1$, and the PHIBSS collaboration}

\affiliation{$^1$Observatoire de Paris, LERMA, College de France, CNRS, PSL, Sorbonne Univ.
UPMC, F-75014, Paris, France \\ email: {\tt francoise.combes@obspm.fr}} 

\pubyear{2015}
\volume{315}  %% insert here IAU Symposium No.
\jname{From interstellar clouds to star-forming galaxies: universal processes?}
\editors{P. Jablonka, P. Andr\'e \& F. van der Tak, eds.}
\begin{document}

\maketitle

\begin{abstract}
The cosmic star formation rate density first increases with time
towards a pronounced peak 10 Gyrs ago (or z=1-2) and then slows down,
dropping by more than a factor 10 since z=1. 
The processes at the origin of the star formation quenching are 
not yet well identified, either the gas is expelled by supernovae and
AGN feedback, or prevented to inflow. Morphological transformation or environment
effects are also invoked.
Recent IRAM/NOEMA and ALMA results are reviewed about the molecular content
of galaxies and its dynamics, as a function of redshift.
Along the main sequence of massive star forming galaxies, the gas fraction
was higher in the past (up to 80\%), and galaxy disks were more unstable 
and more turbulent.
 The star formation efficiency increases with redshift, or equivalently 
 the depletion time decreases, whatever the position of galaxies, either on
the main sequence or above.
Attempts have been made to determine the cosmic evolution of the H$_2$ density,
but deeper ALMA observations are needed to effectively compare with models.
\keywords{Galaxies, Star formation, AGN, Interstellar medium, Molecular gas}
%% add here a maximum of 10 keywords, to be taken form the file $<$Keywords.txt$>$
\end{abstract}

\firstsection % if your document starts with a section,
\section{Introduction}

The cosmic star formation and its evolution through the Hubble time
is now well established (e.g. Madau \& Dickinson 2014). One of the main
issue is to understand the rather fast
winding down of star formation after z=2, both globally and also in
individual galaxies. The suppression of star formation in galaxies appears
rather sudden and therefore is called quenching; this has been
revealed by their bimodal distribution, between a blue cloud of galaxies
actively forming stars, and a red sequence of dead objects. The paucity
of galaxies in an intermediate sequence (e.g. Baldry et al. 2006) means that the
quenching time-scale is typically smaller than 1 Gyr. However the quenching 
mechanism is hard to identify. In a recent study of
56 GOODS galaxies at z=1.7, Mancini et al. (2015) found an equal proportion of AGN
in both star forming and quenched galaxies, and their morphology, traced
by the bulge to disk ratio or their Sersic index, was also comparable over
a large range of specific star formation rate (SFR).

Since star formation is directly linked to the amount of molecular gas
present, and in which state, it is primordial to determine the gas fraction
as a function of redshift, and also the star formation efficiency (SFE),
defined by the ratio between SFR and gas mass.
In the recent years, a large variety of data have been obtained, with
somewhat discrepant results, revealing either low or high SFE at high redshift.
In sub-millimeter galaxies (SMG), and ultra-luminous starbursts (ULIRGs) 
the SFE increases with redshift, and the depletion time (the inverse of SFE)
which is of the order of 0.5-1Gyr for ULIRGs at z$\sim$ 0 becomes as low as 10-100 Myr at z=1-2
(Greve et al. 2005, Combes et al. 2011, 2013). But there exists a population
of massive BzK galaxies, selected from their near-IR and optical colors,
which are also actively forming stars (they are ULIRGs), but with a 
lower efficiency, with a depletion time-scale of the order of 0.3 Gyr
at z$\sim$ 1.5 (Daddi et al. 2008), i.e. similar to local ULIRGs.
 This might be due to their extended molecular component (10 kpc scales),
and their low density state, revealed by their low excitation:
the CO emission as a function of the upper level $J$ peaks at 
$J=3$, as in the Milky Way (Dannerbauer et al 2009), justifiying the
adoption of the standard CO-to-H$_2$ conversion factor (5 times
that adopted for ULIRGs).
 Let us recall that local main-sequence galaxies, such as the
Milky Way, have a depletion time-scale of the order of 2 Gyr 
(Bigiel et al. 2011).

\section{Starburst galaxies}

In a recent work with IRAM and ALMA, 
Silverman et al. (2015) detected CO emission in 7 galaxies from COSMOS, 
with SFR$\sim$ 300-800 M$_\odot$/yr, at 1.4 $<$ z $<$ 1.7.
Their gas fraction was determined to be $\sim$ 30-50\%, and their global SFE is enhanced,
although the objects are unresolved, and the starbursting regions
cannot be separated.

To trace redshift evolution, and in particular the winding down 
of star formation after z=1, the intermediate redshift epoch
0.2 $<$ z $<$ 1 is a key region to explore.
We have observed the gas content of ULIRGs at these intermediate z, 
to make the link with local starbursts (Combes et al. 2011, 2013).
Out of 69 ULIRG, 33 were detected in CO emission, with
variable excitation. Some of the objects have a lower H$_2$ density,
which can be explained in terms of at least two components.
 The objects where an interferometric map is available
show the separation of the molecular emission in a
 nuclear starburst and an extended gas disk of scale $\sim$20 kpc.
The presence of such an extended component may explain lower
SFE, in particular when low gas excitation suggests a higher CO-to-H$_2$
conversion ratio. 

This study at intermediate z, compared to all
other starburst data, showed that both gas fraction
and SFE increase with redshift, by a factor 3$\pm 1$,
between z=0 and 1, with or without taking into account upper limits.

Recently, the Bzk galaxies at z=1.5, which show globally low excitation
of the CO lines at low $J$ were observed to have quite high CO(5-4) fluxes, 
revealing a second component of more excited, denser and warmer molecular gas
(Daddi et al. 2015). For this excited component, the objects have the same 
correlations than for ULIRGs, confirming that the starburst 
occurs either in a separate
nuclear region, or in hot and dense star forming clumps.
In local starbursts, the total CO SLEDs observed with Herschel reveal a very large
variety of shapes, revealing several molecular components (Mashian et al 2015),
suggesting that the overall CO-to-H$_2$ conversion ratio could explore
even wider ranges than expected, from $\alpha$ =0.4 to 5 M$_\odot$/(K km/s pc$^2$)
for the M(H$_2$) to L'CO ratio.

\begin{figure}[t]
\begin{center}
 \includegraphics[width=13cm]{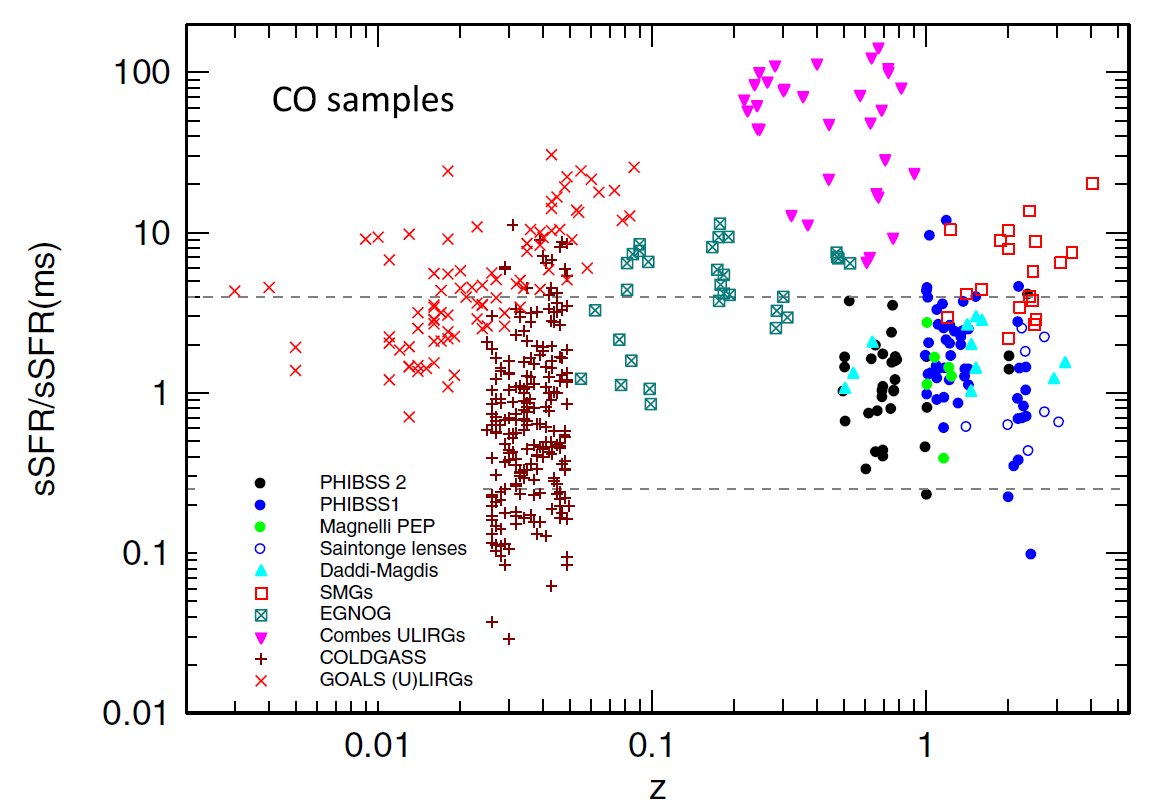} 
% \vspace*{-1.0 cm}
 \caption{Distribution in the redshift versus specific star-formation rate plane 
of all the star forming galaxies with CO flux measurements found in the literature,
compared with the PHIBSS data by Genzel et al. (2015).
The sSFR on the vertical axis is normalized to the main sequence (ms) value
 of sSFR at a given mass and redshift, according to the scaling relation given by
Whitaker et al. (2012). The appartenance to the main sequence is defined
by the two horizontal dashed lines, situated at $\pm$0.6 dex from the mid MS line.}
   \label{fig1}
\end{center}
\end{figure}

\section{Main sequence galaxies}

Most of the cosmic star formation ($\sim$ 90\%) occurs in the main sequence galaxies,
and only about 10\% in starbursts. The main sequence is clearly defined
in an SFR versus stellar mass diagram as a power-law of slope slightly
lower than 1. This power-law is similar at all redshift ranges, but the
zero point is increasing with redshift, following the evolution of
the cosmic star formation rate density described in the introduction.
Large surveys of hundred thousands of galaxies locally (SDSS) or 
at high redshift (GOODS, COSMOS, ..) have shown a 
correlation between morphology and stellar populations since z$\sim$2.5
(Wuyts et al. 2011):
blue star forming galaxies on the main sequence are exponential disks 
(Sersic index near 1), while quiescent red systems are of de Vaucouleurs type
(Sersic index more near 4).

With the goal to explore the molecular gas content of main sequence
galaxies, we undertook the PHIBSS project (Plateau de Bure HIgh z Blue Sequence
Survey, Tacconi et al. 2010, 2013). In the first part of the project,
52 galaxies were detected at IRAM in the CO(3-2) line at z=1.2 and 2.3.
The targets were selected to be massive (M$_* >$ 2.5 10$^{10}$ M$_\odot$)
star forming galaxies (SFR $>$ 30 M$_\odot$/yr). Adopting a standard CO-to-H$_2$
conversion ratio for these main sequence objects, molecular masses were found
between 10$^{10}$ and 3 10$^{11}$ M$_\odot$, corresponding to gas fraction
in average of 33\% at z=1.2 and 47\% at z=2.3. The SFE was found to increase
slightly with z, and the depletion time scale is in average 0.7 Gyr at z=1.2.

\begin{figure}[t]
\begin{center}
 \includegraphics[width=8cm]{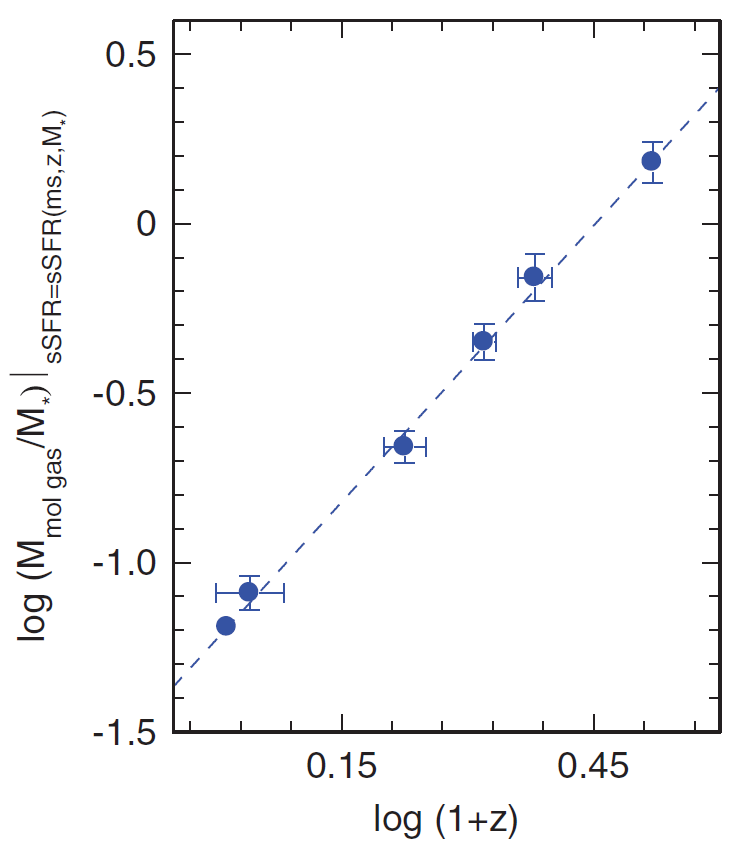} 
 \caption{Ratio of the molecular to stellar mass versus redshift, for galaxies 
on the main-sequence (Genzel et al. 2015). 
The best linear fit has a slope of 2.71 (dashed line). }
   \label{fig2}
\end{center}
\end{figure}

The CO detection rate was quite high ($>$85\%), in these « normal » 
massive Star Forming Galaxies (SFG). Some were mapped at high spatial
resolution, and a rather regular velocity field was found, confirming
the absence of major mergers. At z=1.2, it was possible to resolve four
galaxy disks in clumps with the help of the velocity information, both 
with CO and [OII] lines (gas content and SFR), since there is a
good correlation between molecular and ionised gas. This allowed
us to draw a resolved Kennicutt-Schmidt (KS) relation (Freundlich et al. 2013). 
The high-z points extend the local KS relation towards high gas and SFR surface
densities, with the same slope. For one galaxy at z=1.5, it was possible
to observe the H$\alpha$ line at high resolution from the ground (Genzel et al. 2013).
The KS slope depends strongly on the dust extinction model adopted,
but falls around 1.

\begin{figure}[t]
\begin{center}
 \includegraphics[width=8cm]{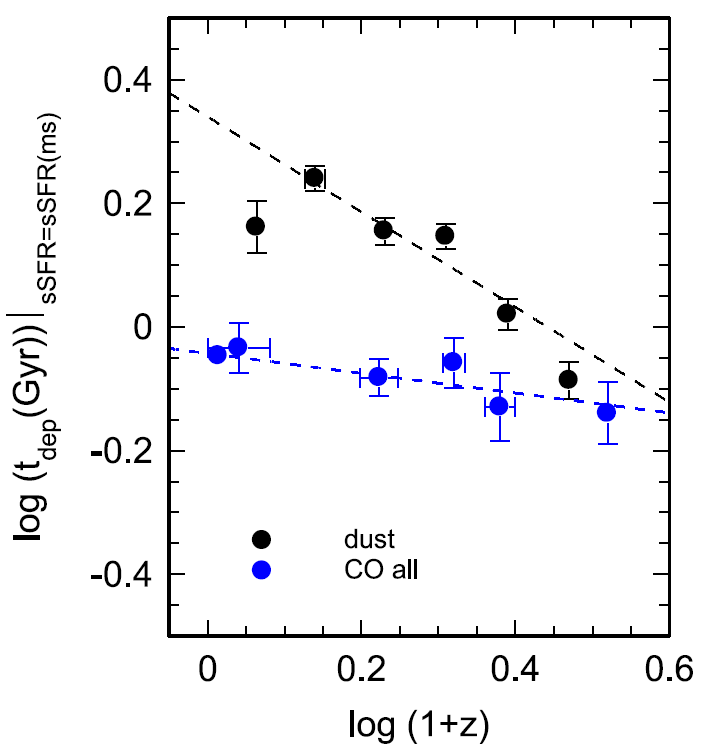} 
 \caption{Depletion time versus redshift for galaxies on the main sequence, 
computed from the dust emission (black circles), and from the CO lines (blue circles).
The best linear fit has a slope of --0.77 (black dashed line).} 
   \label{fig3}
\end{center}
\end{figure}

The evolution of specific SFR with redshift was compatible with 
the results of optical surveys, provided that the depletion time
is varying on the main sequence as 
$t_{dep}$=1.5/(1+z) Gyr.
 The PHIBSS project is now being extended with the goal of observing
CO lines in about 150 galaxies, exploring also intermediate redshifts
(z=0.5-0.7), and some galaxies departing from the main sequence, either starbursts
above the MS or quiescent galaxies below (PHIBSS2). Detecting smaller masses, and 
bright galaxies with more spatial resolution and/or
more molecular lines in addition will be attempted with ALMA.

\begin{figure}[t]
\begin{center}
 \includegraphics[width=13cm]{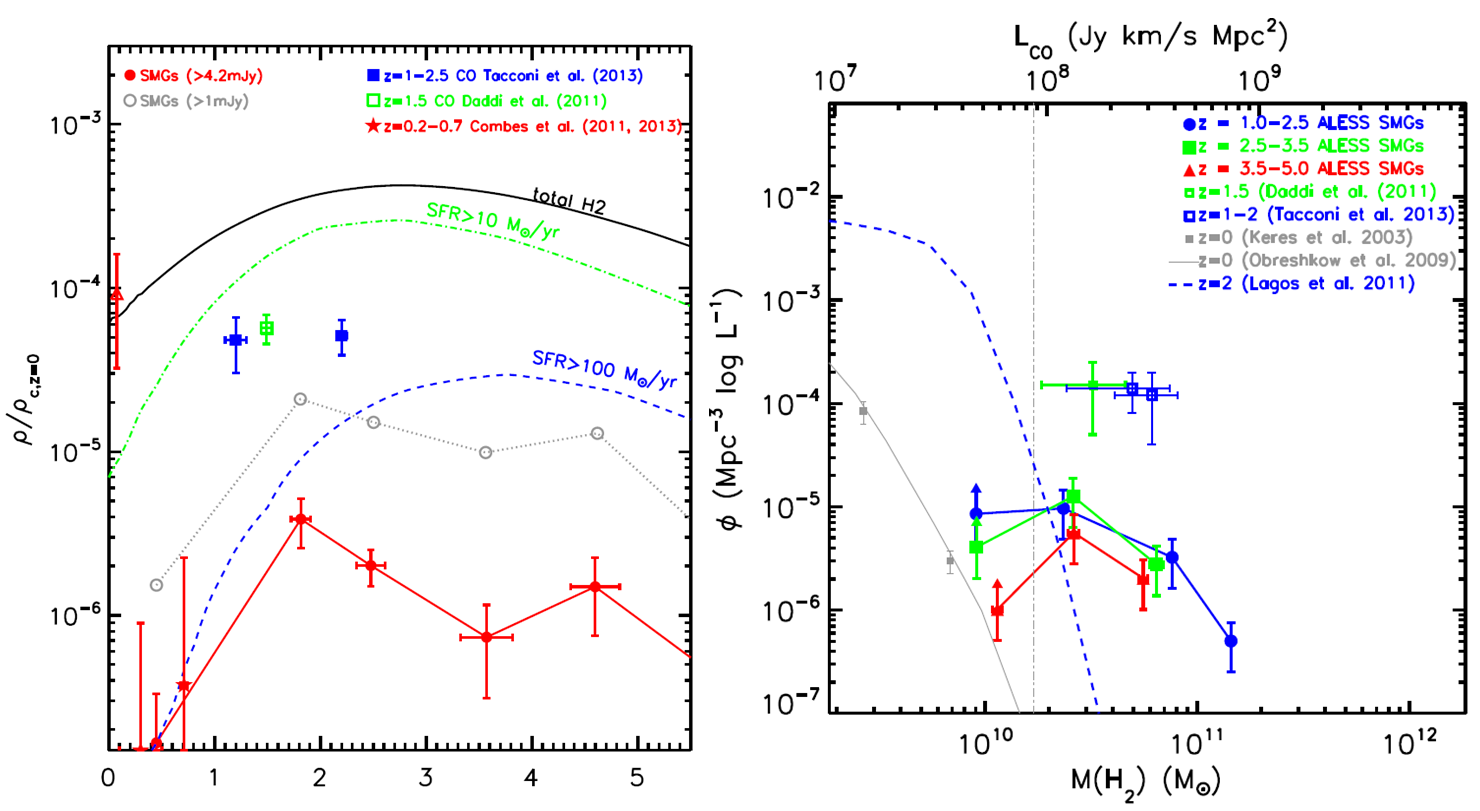} 
 \caption{ {\bf Left:}  The normalised H$_2$ density $\Omega$(H$_2$) as a function 
of redshift, for SMG and other high-z starbursts 
(Swinbank et al 2014, Combes et al. 2013) in
red, compared to main sequence galaxies in blue (Tacconi et al 2013). The dashed lines
are the model predictions, for different SFR (Lagos et al 2011).
{\bf Right:} The H$_2$ mass function for SMG and main sequence galaxies
compared with models. The main sequence galaxies at z=1.5 (green) are from Daddi et al. 
(2010) and at z=1-2 (blue) from Tacconi et al. (2013). They are well above the 
predictions at z=2 by Lagos et al (2011) indicated by the blue dashed curve. 
 From Swinbank et al. (2014).}
   \label{fig4}
\end{center}
\end{figure}

The first results and the scaling relations obtained on 
the main sequence, by comparison with all other data with CO detections at high redshift
 (cf Fig \ref{fig1}), were presented in Genzel et al. (2015).
On the MS, the gas fraction increases regularly with z, as shown in Fig \ref{fig2}.

Fig \ref{fig3} displays the depletion time as a function of redshift,
for the galaxies on the main sequence. There is a slight decrease,
and the effect depends on the way the molecular gas content is estimated,
either from the CO line, or from the dust emission.

A recent survey with ALMA of the continuum dust emission of 180 star forming
galaxies between z=1 and 6.4 results in slightly different results
(Scoville et al. 2015). The gas fraction estimated from the dust emission, assuming 
a constant dust temperature, is also highly increasing with redshift, reaching
values as high as 50-80\%. The depletion time-scale is found to be the same
for starbursts and main sequence galaxies; it decreases strongly with redshift,
to reach 200 Myr at z$>1$, therefore 10 times lower than for local MS galaxies.

\section{The cosmic H$_2$ density}

One of the key issues to understand the cosmic star formation history
is to observe the cosmic evolution of the H$_2$ density.
Theoretical considerations and semi-analytical models predict that the
molecular gas density must increase with redshift, and dominate over the
atomic gas in galaxies (Obreschkow \& Rawlings, 2009, Obreschkow et al. 2009). 
The phase transition from atomic to molecular hydrogen
is favored by pressure (Blitz \& Rosolowsky 2006),  while
surface density and consequently the pressure is higher in high-z galaxies. 
The modelisation predicts a molecular-to-atomic ratio
 H$_2$/HI varying as (1+z)$^m$, with m as high as 1.6.
Some models however predict flatter evolutions of the
H$_2$ density at high z (Lagos et al. 2011, Popping et al. 2014).

 Decarli et al. (2014) and Walter et al. (2014) have attempted to constrain 
the H$_2$ density by observing a large cosmic volume of $\sim$7000 Mpc$^3$, in 
the Hubble deep field North with Plateau de Bure. They 
separate the results in three redshift bins: z$<$0.45, 1.01$<$z$<$1.89 and  z$>$2.
A blind molecular line survey has been carried out through scanning the whole 3mm
band. The blind detection of 17 CO lines, 
together with the upper limits obtained by stacking the observations towards 
spectroscopically identified
objects, constrain the CO luminosity functions at the corresponding redshifts. 
 The results show that optical/mid-IR bright galaxies contribute less 
than 50\% to the star formation rate density at 1 $<$ z $<$ 3, and the normalised
density $\Omega$(H$_2$) 
at high z tends to be higher than the predictions.

It might be easier and certainly quicker to determine the evolution of the H$_2$ density
from dust emission surveys.
A recent 870$\mu$m continuum survey with ALMA of 99 SMG in the Extended Chandra
Deep Field South (Swinbank et al. 2014) has discovered that the well detected
sources (S$_{870}$ $>$ 4.2 mJy) are in average ULIRGs with SFR=300 M$_\odot$/yr,
and dust temperatures of 32 K. They contribute to only 1-2\% of SFR. 
The extrapolation of the counts down to S$_{870} >$ 1 mJy 
through stacking shows that these sources contribute to 20\% of
the cosmic star formation density over z=1-4 (see Fig \ref{fig4}). 
Deriving H$_2$ masses from dust masses,
the average SFE is found rather high, with depletion time-scale of 130 Myr.

\section{Conclusions}

It is now well established that galaxies at high redshift
 have a larger gas fraction than local ones,
whatever their position on the main sequence or above,
in the starburst domain. The gas fraction can reach 50\% and above.

There is not yet a consensus on the exact evolution of
the star formation efficiency with redshift. The inverse of the SFE,
the depletion time scale, is decreasing with redshift, however the
amplitude of its variation with z is still debated.  The various results
depend on the way to estimate the total molecular gas amount,
either from CO lines or from dust emission.
The results may also depend on 
 the definition of the Main Sequence (e.g. Renzini \& Peng, 2015).

 A higher SFE at high z might be explained by a higher surface density
of molecular gas, if the Kennicutt-Schmidt relation is non-linear, which is
not yet well known. Alternatively, a
starburst can be triggered in nuclear regions when the gas is concentrated.
Diagnostics could be searched for with CO excitation and the observation
of several $J$ lines, and also dense gas tracers (HCN, HCO$^+$).

ALMA observations begin to estimate the evolution
of the molecular gas mass in galaxies, however we are still far from a 
total census of $\Omega$(H$_2$) as a function of redshift.

\end{document}